\documentclass[aps,prl,groupedaddress,superscriptaddress,twocolumn,showpacs]{revtex4-1}

\usepackage{graphicx}
\usepackage{dcolumn}
\usepackage{bm}
\usepackage[T1]{fontenc}
\usepackage[french]{babel}
\usepackage{epstopdf}
\usepackage{amssymb}
\usepackage{color}
\usepackage{amsmath}
\usepackage{subfigure}
\usepackage{natbib}

\begin{document}

\preprint{APS/123-QED}

\title{Transport through graphene on SrTiO$_3$}

\author{Nuno~J.~G.~Couto}
\author{Benjamin~Sacépé}
\author{Alberto~F.~Morpurgo}
\affiliation{
Départment de Physique de la Matière Condensée (DPMC) and Group of Applied Physics (GAP), University of Geneva, 24 Quai Ernest-Ansermet 1211 Genéve 4, Switzerland
}%

\begin{abstract}
We report transport measurements through graphene on SrTiO$_3$ substrates as a function of magnetic field $B$, carrier density $n$, and temperature $T$. The large dielectric constant of SrTiO$_3$ screens very effectively long-range electron-electron interactions and potential fluctuations, making Dirac electrons in graphene virtually non-interacting. The absence of interactions results in a unexpected behavior of the longitudinal resistance in the $N=0$ Landau level, and in a large suppression of the transport gap in nano-ribbons. The "bulk" transport properties of graphene at $B=0$ T, on the contrary, are completely unaffected by the substrate dielectric constant.
\end{abstract}

\maketitle

Experiments on devices with SiO$_2$\cite{Peres_RevModPhys.82.2673} and BN\cite{Dean_Kim_Graphene_in_hBN_NatNano} gate dielectrics, as well as on suspended layers\cite{KimSolidStateComm,*Andrei_Ballistic_Suspended_NatNano}, indicate that the substrate material has a strong influence on the transport properties of graphene. Whereas investigations have mainly aimed at minimizing the amount of disorder present, it should be possible to choose the substrate material to effectively control different aspects of the electronic properties of graphene. Here we discuss transport experiments through graphene on SrTiO$_3$, a very well-known insulator where the presence of a low-energy phonon mode\cite{WorlockPR1968,*MullerPRB1979} brings the material close to a ferroelectric instability. The softening of this low-energy mode causes the relative dielectric constant ($\epsilon$) of the material to increase from 300 at room temperature, to $\simeq 5000$  at liquid Helium temperature\cite{MullerPRB1979}, with most of the increase taking place when $T$ is lowered from 50 to 10 K. The use of SrTiO$_3$ substrates, therefore, allows the investigation of the effect of the dielectric environment on the charge carriers in graphene in a broad temperature range, influencing all phenomena in which electron-electron interactions and long-range electrostatic potentials play a dominant role.

The effect of dielectric screening in graphene has been investigated previously, mainly focusing on its effects on carrier mobility. In one set of experiments, graphene devices were immersed in solvents with $\epsilon$ up to $50-100$\cite{GeimPRL102_solvents}. Although revealing, these studies have been confined to temperatures above the solvent freezing point ($T> 160$ K), which prevented the investigation of physical phenomena taking place at low temperature. In other work, the effect of the dielectric environment was investigated by comparing the transport properties of graphene on SiO$_2$ with and without an adsorbed thin ice layer which slightly changed the dielectric environment seen by the charge carriers\cite{Fuhrer_Ice_layers_PhysRevLett.101.146805}. The use of SrTiO$_3$ offers the advantage of a very large tunable dielectric constant together with the possibility of measuring in a broad range of temperatures.

Our studies rely on transport measurements on graphene Hall-bars and etched nano-ribbons on SrTiO$_3$. At zero magnetic field, transport through Hall-bar devices show no temperature dependence between 250 mK and 50 K, and graphene exhibits a behavior identical to that observed on SiO$_2$ substrates\cite{Fuhrer_PRB_Charge_and_inhomo}. The importance of the  substrate, however, becomes clear in measurements at finite $B$, and in nano-ribbons. In the quantum Hall regime, we observe that the longitudinal resistance peak measured in the $N=0$ Landau level decreases significantly with lowering $T$, a trend opposite to what is seen on common SiO$_2$ substrates\cite{OngPRL2008,*Maan_Zeitler_Gap_opening_PhysRevB.80.201403,*Zaliznyak2010}. In nano-ribbons, the magnitude of the transport gap is one order of magnitude smaller than in identical devices on SiO$_2$\cite{Kim_nano_rib_PhysRevLett.98.206805,*Oostinga2010,*Ensslin2010,*Goldhaber-Gordon2009}. Both effects are manifestations of the suppression of electron-electron interactions due to substrate screening, which turns carriers in graphene into virtually non-interacting Dirac fermions. The observation of such a very effective screening also allows us to conclude that at $B=0$ T Coulomb interactions and long-range potentials do not have any significant influence on transport through "bulk" graphene on SiO$_2$.

\begin{figure}[h!]
\begin{center}
\includegraphics[width=0.9\linewidth]{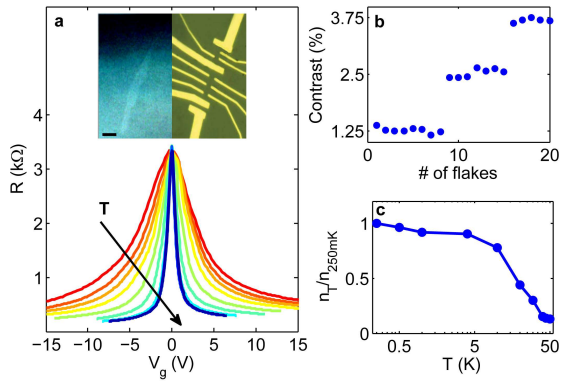}
\caption{(Color online) (a) Square resistance of graphene on SrTiO$_3$ as a function of gate voltage, measured for different temperatures between 50K and 250mK; the arrow points in the direction of lowering temperature. (a-inset) Optical microscope image with software enhanced contrast(left); the final device with contacts attached (right; scale bar is $3\mu m$ long). (b) Optical contrast of mono-, bi-, and tri-layer graphene on SrTiO$_3$.  (d) Ratio of the carrier density measured by Hall effect at V$_g=2$ V at temperature $T$ over the carrier density at $T=250$ mK: this ratio is proportional to the dielectric constant of the substrate $\epsilon(T)$, and decreases by one order of magnitude when $T$ is increased from 250 mK to 50 K, as expected.}
\end{center}
\label{FIG.1}
\end{figure}

The devices investigated (Fig. 1(a)-inset) were prepared by exfoliating graphene from natural graphite using an adhesive tape, and by the subsequent transfer onto a 500$\mu$m thick SrTiO$_3$ single crystalline substrate. Graphene layers were found by inspection under an optical microscope, with mono-, bi-, and tri-layer graphene giving a contrast of  $1.25\%$, $2.5\%$ and $3.75\%$, respectively (see Fig. 1.(b); this contrast is measured on substrates with only one face polished). Ti/Au contacts (10/50 nm) were defined by means of electron beam lithography, evaporation, and lift-off (a thin Cu layer was evaporated onto the PMMA prior to performing lithography to prevent surface charging, and was etched away with a FeCl solution before development). The gate electrode consisted of a Au film evaporated on the substrate backside.

We first discuss measurements performed  as a function of gate voltage  $V_g$ and at $B=0$ T, at  temperatures between 50 K and 250 mK. The increase in dielectric constant with lowering $T$ is clearly visible in Fig. 1.(a), where the graphene resistance measured at each temperature is plotted as a function of $V_g$. With lowering $T$, a smaller  $V_g$ range is needed to scan across the resistance peak  around the charge neutrality point ("Dirac peak"). This is a direct consequence of the increase of the gate capacitance, due to the increase in the substrate dielectric constant. At each gate voltage, the density of charge carriers $n$ was extracted from the Hall resistance, and Fig. 1(c) shows that at a fixed value of $V_g$,  $n$ increases by approximately one order of magnitude as $T$ is lowered from 50K down to 250mK, in agreement with the expected behavior of  SrTiO$_3$. Fig. 2(a) shows that, when plotted as a function of  $n$, the Dirac peaks measured at all different temperatures overlap nearly perfectly despite the tenfold change in $\epsilon$ of the substrate. This result has several direct implications that will be discussed in detail later.

\begin{figure}[h]
\begin{center}
\includegraphics[width=0.9\linewidth]{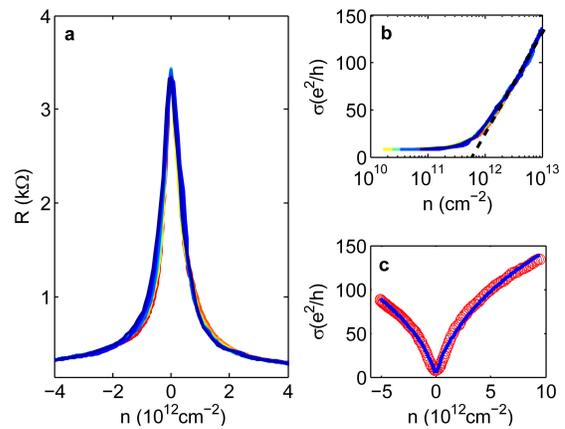}
\caption{\label{fig:fig2} (Color online)(a) Square resistance of graphene on SrTiO$_3$ measured at different temperatures between 250 mK and 50 K,  as a function of carrier density (extracted from Hall effect measurements). (b) Conductivity $\sigma$ of graphene on SrTiO$_3$  as a function of $n$ (in log scale) at different temperatures, showing the low density region where  $\sigma$ is independent of $n$ (by extrapolating the $\sigma(n)$ curve measure at large density --dotted line-- we estimate the width of this region to be approximately $\delta n=(6.0 \pm 0.5) \times 10^{11} \textrm{cm}^{-2}$, independent of temperature).(c -blue line) Fitting the experimental data to the dependence of $\sigma (n)$  by taking into account resonant scattering  (eq.(1) of main text) gives a very good agreement with values for the parameters ($n_i= 2.9\times 10^{11}\textrm{cm}^{-2}$  $R=0.22$nm) very close to those obtained for graphene on SiO$_2$.}
\end{center}
\end{figure}

Finding a complete insensitivity to temperature despite the large change in the SrTiO$_3$ dielectric constant may cause doubts that water or other adsorbed molecules are present in between the graphene layer and the substrate, effectively decoupling the two materials. To rule out this possibility, it is important to identify measurable effects sensitive to the substrate dielectric constant. Obvious candidates are phenomena that originate from  long-range electron-electron interaction in graphene, which should be completely screened on SrTiO$_3$. Two such phenomena are the temperature dependence of the longitudinal resistance in the $N=0$ Landau level in the quantum Hall regime\cite{OngPRL2008,*Maan_Zeitler_Gap_opening_PhysRevB.80.201403,*Zaliznyak2010}, and low-temperature bias-dependent transport in nano-ribbons\cite{Guinea_PhysRevLett.99.166803,Kim_nano_rib_PhysRevLett.98.206805,*Oostinga2010,*Ensslin2010,*Goldhaber-Gordon2009}.

Figure 3 shows data measured in the presence of a 15 $T$ perpendicular magnetic field. At high magnetic field, well-defined Hall plateaus in the Hall conductivity $\sigma_{xy}$ are observed up to the maximum temperature investigated (50 K; see Fig. 3b). The plateaus  at $\sigma_{xy} = \pm 2$ and $\sigma_{xy} = \pm 6 e^2/h$  confirm that the device is of good quality and that it indeed consists of a single graphene layer\cite{Geim_2D_diracFermion_gas,*Kim_Exp_QHE_Graphene_Nature}. Particularly interesting is the temperature dependence of the peak in the longitudinal resistance  observed at the charge neutrality, when the Fermi level is located inside the $N=0$ Landau level, with the height of this peak decreasing monotonously with lowering temperature from 50 K to 250 mK. This behavior is the opposite of what has been by now reported by several groups for graphene on SiO$_2$\cite{OngPRL2008,Zaliznyak2010,Maan_Zeitler_Gap_opening_PhysRevB.80.201403}, or for suspended graphene\cite{KimSG_FQHE_Nat,*andreiSG_FQHE_Nat}. In these cases, an insulating behavior is observed, accompanied by a very rapid (thermally activated below $T \approx 20$ K) increase in the resistance at charge neutrality with lowering $T$. The insulating behavior at $N=0$ is attributed to a symmetry broken state due to electron-electron interactions\cite{Peres_Elec_Prop_Disor_PhysRevB.73.125411,*Nomura_Macdonald_PhysRevLett.96.256602}. The observation of a decrease in resistance with lowering temperature shows that the symmetry broken state is absent in graphene on SrTiO$_3$. This observation substantiates that the insulating behavior at $N=0$ is indeed due to Coulomb interactions, and provides a direct manifestation of the effective substrate screening.

\begin{figure}[h]
\begin{center}
\includegraphics[width=1\linewidth]{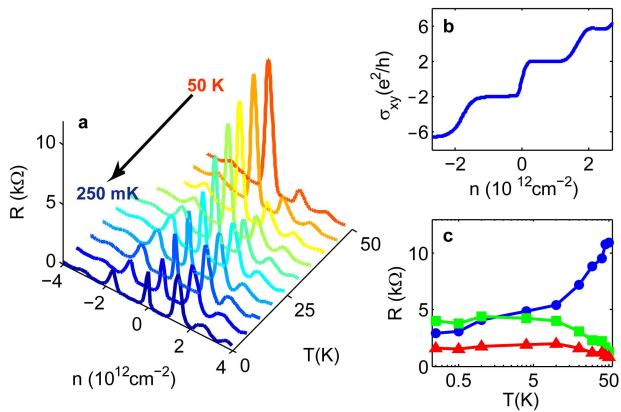}
\caption{\label{fig:fig3} (Color online)(a) Longitudinal square resistance $R$ as a function of V$_g$ at B =15 T, measured at different temperatures between 250 mK and 50 K. The height of the resistance peak at the charge neutrality ($N= 0$ Landau level) decreases significantly with lowering $T$, whereas the height of the peaks centered around the subsequent $N= \pm 1$ Landau shows a slight increase in resistance with lowering $T$ (see panel (c); $N=0$(circles), $N=+1$(squares) and $N=  -1$(triangles)).  (b) The Hall conductivity measured at 50 K (B=15 T) shows very well developed plateaus at $\pm 2$, $\pm 6$ $e^2/h$, as it characteristic for Dirac fermions in graphene.}
\end{center}
\end{figure}

The effect of substrate screening is also visible in  bias-dependent transport measurements on etched nano-ribbons. When these nano-ribbons are gate-biased near the charge neutrality point, a transport gap opens, due to the joint effect of disorder and Coulomb interactions\cite{Kim_nano_rib_PhysRevLett.98.206805,Ensslin2010,Goldhaber-Gordon2009,Oostinga2010}. Disorder causes electrons to localize in small regions of the ribbons, making charging effects important. In simple terms, a graphene nano-ribbon behaves as an array of Coulomb-blockaded quantum dots, and the transport gap originates from the charging energy of the individual islands\cite{Guinea_PhysRevLett.99.166803}. When the bias applied across the ribbon is sufficient to overcome the Coulomb gap, the differential conductance increases, providing the means to measure the gap magnitude. On devices realized on SiO$_2$ substrates, it has been found that the largest gap in source-drain voltage scales approximately inversely to the ribbon width $W$, and for a 1 $\mu$m long and  70 nm wide ribbon, it is approximately 15-20 meV\cite{Oostinga2010,Ensslin2010}. Fig. 4(a) shows the conductance of a 70 nm wide ribbon on SrTiO$_3$ (approximately 1 $\mu$m long) as a function of gate and bias voltage ($V_{sd}$). It is apparent that the conductance is suppressed at low bias, and the data appear qualitatively similar to those routinely measured on samples on SiO$_2$. A closer look reveals that the largest size of the gap is only 2 meV, i.e. one order of magnitude less than for identical devices SiO$_2$. The size of the gap is more clearly seen in Fig. 4(b), which shows the bias dependence of the conductance measured for three different values of gate voltages. The large reduction of the transport gap in SrTiO$_3$ (as compared to SiO$_2$) devices originates from the large substrate dielectric constant, which strongly increases the self-capacitance of the islands in the ribbon, thereby suppressing the charging energy (note that the remaining gap of 2 meV also includes a contribution due to single particle level spacing, which is present even if the charging energy is completely suppressed, and that is likely to be dominant for ribbons on SrTiO$_3$). The observation of a strongly suppressed gap in nano-ribbons on SrTiO$_3$, therefore, provides a second clear indication of the effectiveness of substrate screening.

\begin{figure}[h]
\begin{center}
\includegraphics[width=0.9\linewidth]{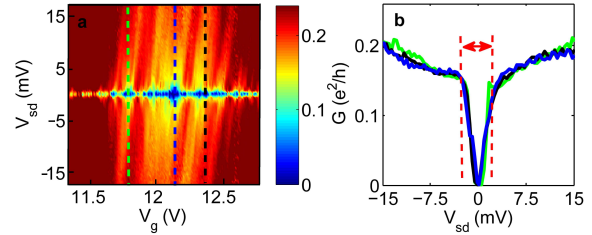}
\caption{\label{fig:fig4}(Color online)(a) Differential conductance $G$(in units of $e^2/h$), of a graphene nano-ribbon, as a function of the source-drain and gate voltage, $V_{sd}$ and $V_g$. (b) Differential conductance as a function of $V_{sd}$ for three specific $V_g$ values, corresponding to the blue, black and green dotted lines in panel (a). The interval enclosed by the red dashed lines gives the size of the transport gap, which is approximately 2 meV for each polarity of $V_{sd}$ (one order of magnitude smaller than for an identical ribbon on SiO$_2$).}
\end{center}
\end{figure}

Having established that substrate screening is effective, we go back to discuss the implications of the behavior observed at $B=0$ T, where transport measurements are found to be completely insensitive to temperature. From these measurements we directly conclude that neither the carrier mobility ($\mu \simeq 7000$ cm$^2$/Vs at $n=2\times 10^{12}\textrm{ cm}^{-2}$, decreasing slightly with increasing density) nor the width $\delta n$ of the low density region where the conductivity $\sigma$ is independent of $n$ (see Fig.2b; $\delta n \simeq 6.0 \pm 0.5 \times 10 ^{11} \textrm{ cm} ^{-2}$) are affected by the large change in the dielectric constant of the substrate. Additionally, the measured values of $\mu$ and $\delta n$ are essentially identical to those measured on typical graphene devices on SiO$_2$\cite{Fuhrer_PRB_Charge_and_inhomo}, where the dielectric constant is roughly 1000 times smaller. These findings directly rule out long-range Coulomb potentials generated by charge impurities next to the graphene layers as the main, mobility-limiting scattering mechanism in graphene\cite{Das_sarma_CArrier_2d_graphene_PhysRevLett.98.186806,*Das_sarma_PNAS,*Fuhrer_NatPhy_Charged_imp} (in agreement with the conclusions in Ref.\cite{GeimPRL102_solvents}, which only explored the high temperature regime). Our experimental results are better explained by  resonant scattering due to impurities generating very strong and short range potentials \cite{StauberPRB,*KatsnelsonPRL2010,*CastroNetoPRB2011}, which lead to the following expression for the carrier density dependence of the conductivity $\sigma$:
\begin{eqnarray}
\sigma = \frac{2e^2}{\pi h}\frac{n}{n_i}\ln^2(\sqrt{n\pi}R)
\end{eqnarray}
($n_i$, is the impurity concentration, and $R$ the potential range respectively). Fig. 2(c) shows that this expression provides a satisfactory fit to our data, with values for the parameters $n_i = 2.9\times 10^{11} \textrm{cm}^{-2}$ and $R=0.22$nm that are very close to those found for graphene on SiO$_2$\cite{Fuhrer_Defect_scattering_PhysRevLett.102.236805,*Monteverde_PhysRevLett.104.126801}. Indeed,  the potentials responsible for resonant scattering have very short range and cannot be screened by the substrate, which explains why the $B=0$ T conductivity and mobility of carriers in graphene on SrTiO$_3$ and on SiO$_2$ are the same\footnote{Another scattering mechanism that is compatible with our observations is scattering by ripples\cite{Geim_Katsnelson_corrugat,Peres_RevModPhys.82.2673}, which leads to a density dependence of the conductivity similar to that of resonant scatterers and is not affected by the dielectric environment}.

Regarding  $\delta n$, it is often assumed that this quantity is determined by potential fluctuations which give origin to the so-called "puddles", i.e. spatial fluctuation in carrier density whose presence has been observed in different kinds of experiments\cite{Yacoby_puddles_NatPhys,*Crommie_STM_NatPhy,*Leroy_STM_PhysRevB.83.155409,Fuhrer_Kelvin_probe_APL}. Our results clearly indicate that the density fluctuations that affect the conductivity cannot be long-ranged: on SrTiO$_3$ all long-ranged potential and charge density fluctuations are very strongly suppressed as compared to SiO$_2$, and still the value of  $\delta n$ on SrTiO$_3$  and on SiO$_2$ coincide. Consistently with this conclusion, $\delta n$ remains temperature independent despite the large increase in $\epsilon$ with lowering $T$ that occurs in the investigated temperature range. In contrast to long-range fluctuations, fluctuations of carrier density that vary on a length scale of at most a few nanometers (not much larger than the graphene-substrate distance) cannot be screened effectively by the substrate and can account for our observations. Indeed, density fluctuations on such a length scale are present in graphene on SiO$_2$, as it has been revealed experimentally by scanning tunneling experiments\cite{Crommie_STM_NatPhy}.

Finally, from the temperature independence of the resistance peak at the charge neutrality point we also conclude that scattering from phonons in the substrate\cite{Fratini_Guinea_phonon_PhysRevB.77.195415} does not play an important role for graphene on SrTiO$_3$. For devices on SiO$_2$, the resistance temperature dependence becomes appreciable above $\simeq 150$ K, and its origin has been attributed to scattering from phonons in the substrate, whose relevant energy in SiO$_2$ is approximately 60 meV\cite{FuhrerNatureNano_intrinsic_limits_G_on_SiO2}. In SrTiO$_3$, phonons with very low energy --down to 3-4 meV at low temperature-- are present, which should give a sizable temperature dependence of the resistance already starting from 10-15 K\cite{WorlockPR1968}. However, no temperature dependence is observed throughout the temperature range investigated (up to 50 K).

In summary, we have performed a complete study of low-temperature transport through graphene on SrTiO$_3$ which, through a comparison with results obtained on lower dielectric constant substrates, enables a direct identification of phenomena that originate from long-range electron-electron interactions and Coulomb potentials. Whereas on low dielectric constant substrates, interactions play a crucial role in determining the properties of the $N=0$ Landau level and the size of the transport gap in nano-ribbons, in graphene on SrTiO$_3$ they are very strongly suppressed by substrate screening. In the latter case, charge carriers in graphene can be thought of a two-dimensional non-interacting gas of Dirac electrons. Since at $B=0$ T no difference in the transport properties of graphene on SiO$_2$ and SrTiO$_3$ is found, we can also conclude that long-range  potentials play virtually no role in determining the ($B=0$ T) density-dependent conductivity of "bulk" graphene on common substrates.

We acknowledge A. Ferreira for technical assistance, A. Caviglia, M. Fogler, S. Gariglio, I. Martin, J. B. Oostinga and N. M. R. Peres for discussions, SNCF and the NCCRs MANEP and QSIT for financial support.

\bibliography{Ref_List_Transport_through_graphene_on_SrTiO3}

\end{document}